\documentclass[twocolumn,pre,showpacs,superscriptaddress,floatfix]{revtex4}
\usepackage{graphicx}
\usepackage{amsmath}
\usepackage{amssymb}
\usepackage{bm}
\usepackage{color}
\DeclareMathOperator\erfc{erfc}

\begin{document}
\title{Optimal tuning of a confined Brownian  information engine}

\author{Jong-Min Park}
\affiliation{Department of Physics, University of Seoul, Seoul 130-743, Korea}
\author{Jae Sung Lee}
\affiliation{School of Physics, Korea Institute for Advanced Study,
Seoul 130-722, Korea}
\author{Jae Dong Noh}
\affiliation{Department of Physics, University of Seoul, Seoul 130-743, Korea}
\affiliation{School of Physics, Korea Institute for Advanced Study,
Seoul 130-722, Korea}

\date{\today}

\begin{abstract}
A Brownian information engine is a device extracting a mechanical work 
from a single heat bath by exploiting the information on the state of a 
Brownian particle immersed in the bath.
As for engines, it is important to find the optimal
operating condition that yields the maximum extracted work or power.
The optimal condition for a Brownian information engine with a finite cycle
time $\tau$ has been rarely studied because of the difficulty in finding the
nonequilibrium steady state.
In this study, we introduce a model for the Brownian information engine and
develop an analytic formalism for its steady state distribution for any $\tau$. 
We find that the extracted work per engine cycle is maximum when $\tau$ 
approaches infinity, while the power is maximum when $\tau$ approaches zero. 
\end{abstract}


\maketitle

\section{Introduction}

The information engine refers to a system extracting a work from a single
heat bath by using the information on the microscopic state of the system.
Discussions on the information engine date back to the thought experiment
on Maxwell's Demon suggested in 1871~\cite{{Leff:2003md}}.
Through the thought experiment, Maxwell claimed that the entropy can
be decreased apparently by performing measurements and feedback controls on
a thermodynamic system. 
Later on, Szilard~\cite{Szilard:1929se} proposed a primary model for the
information engine. In this model he showed that a work can be extracted from
a single heat bath, the entropy of which decreases.
These examples had been regarded as a paradox because the thermodynamic
second law prohibits the total entropy from decreasing. However, in 2009,
Sagawa \emph{et al.}~\cite{Sagawa:2009it} resolved this paradox by
discovering the information fluctuation theorems
~\cite{{Sagawa:2009it},{Sagawa:2010th}}; they showed that the thermodynamic 
entropy (work) can be decreased (extracted) as much as the mutual information
gain by the measurement.
After this discovery, there has been a surge of interest in studying the information fluctuation theorems~\cite{{Ponmurugan:2010th},{Horowitz:2010th},{Sagawa:2011th},{Vaikuntanathan:2011th},{Sagawa:2012th},{Sagawa:2012th2},{Lahiri:2012th},{Hartich:2014th}} and developing theoretical models for the information engine from classical~\cite{{Mandal:2012},{Barato:2014},{Horowitz:2011op},{Bergli:2014op},{Pal:2014op},{Abreu:2011},{Abreu:2012th},{Mandal:2013di},{Bauer:2012},{Kosugi:2013}} to quantum~\cite{{Kim:2011qe}} systems.
With the help of technological advancement, several information engines have
been realized in electronic~\cite{{Koski:2014ex},{Koski:2014ex2}} and
Brownian systems~\cite{{Lopez:2008ex},{Toyabe:2010ex}}.

Among many examples, Brownian systems are a good test base for the
classical stochastic theory based on the Langevin or Fokker-Plank equations.
For this reason, many researchers have studied the information engines
consisting of a Brownian particles trapped in a harmonic
potential~\cite{{Pal:2014op},{Abreu:2011},{Bauer:2012},{Kosugi:2013}}. For
example, Abreu \emph{et al.}~\cite{Abreu:2011} studied the case
where the potential center is varied, Bauer \emph{et al.}~\cite{Bauer:2012} 
studied the case where the potential center and the stiffness are
varied, and Kosugi~\cite{Kosugi:2013} investigated a similar problem. 

In a practical aspect, the primary concern for the Brownian information 
engine lies in the efficiency.
More specifically, we are interested in two quantities: 
The extracted work per engine cycle and the extracted work per unit time, 
i.e., the power. 
In a classical heat engine without exploiting any information, the maximum
efficiency is achieved when the engine is operated quasi-statically and
reversibly. However, the power vanishes in a reversible engine and the
condition for the maximum power is different from that for the maximum
efficiency~\cite{Curzon:1975}. 
In this work, we investigate the optimal condition
for the extracted work per engine cycle or the power in a model for the 
Brownian information engine.

In spite of its practical importance, the optimal tuning of the Brownian
information engine has been studied rarely due to 
the difficulty in finding the nonequilibrium steady state of an
engine having finite engine cycle time $\tau$. The optimal tuning 
conditions have been studied mostly for engines
with $\tau=\infty$~\cite{Abreu:2011}. Kosugi~\cite{Kosugi:2013}
developed a formalism for finite $\tau$, but only the
infinite $\tau$ limit was addressed. 

In this study, we introduce a model for the information engine consisting of a
Brownian particle confined in a harmonic potential. In this model, one
engine cycle of duration $\tau$ consists of the three processes:
measurement of the particle position, feedback control of the potential
center, and relaxation. We derive a self-consistent equation for the steady
state probability distribution function for general $\tau$, whose solution
is found in a series expansion form. 
Using this formalism, we obtain the optimal parameters set that yields
the maximum extracted work per cycle and the maximum power. 
We find that the global maximum of the extracted work per
cycle is realized when $\tau$ is taken to be infinity. On the other hand, the
global maximum of the power is achieved in the $\tau \to 0 $ limit.

This paper is organized as follows. In Sec.~\ref{sect_model}, 
we introduce our model. 
In Sec.~III, we develop a formalism for the nonequilibrium steady state
distribution of the system. Using the formalism, we investigate the optimal
condition for the maximum work per cycle and the power in Sec.~IV. 
We conclude the paper with summary in Sec.~V.

\section{Description of the Model}\label{sect_model}

We consider a one-dimensional overdamped Langevin dynamics of a Brownian
particle in a heat bath with temperature $T$. The particle is 
confined by an external harmonic potential $V(X,\lambda(t)) =
\frac{1}{2}k(X-\lambda(t))^2$ where $X$ is the position of the Brownian
particle, $k$ is a stiffness constant, and $\lambda(t)$ denotes a 
time-dependent potential center with $\lambda(0)=0$. 
This dynamics is described by the Langevin equation
\begin{equation}\label{Langevin_eq}
\gamma \frac{dX}{dt}  =   -k(X-\lambda(t))+\xi(t), 
\end{equation}
where 
$\gamma$ is the damping coefficient, and $\xi \left( t \right)$ is 
a Gaussian white noise satisfying $\left < \xi \left( t \right) \right > = 0$ and 
$\left < \xi \left( t \right) \xi \left( t^\prime \right) \right> = 2 
{\gamma k_B T} \delta \left(t-t^\prime \right)$ with the Boltzmann constant
$k_B$.  The bracket $\left< \cdots \right>$ means the ensemble average. Note
that the Langevin dynamics (\ref{Langevin_eq}), when $\lambda(t)$ is 
time-independent, is known
as the Ornstein-Uhlenbeck process~\cite{Gardiner:2010tp, Risken:1996vl}.
For convenience, we will set $\gamma=k=k_B T=1$. 

\begin{figure}
\includegraphics*[width=\columnwidth]{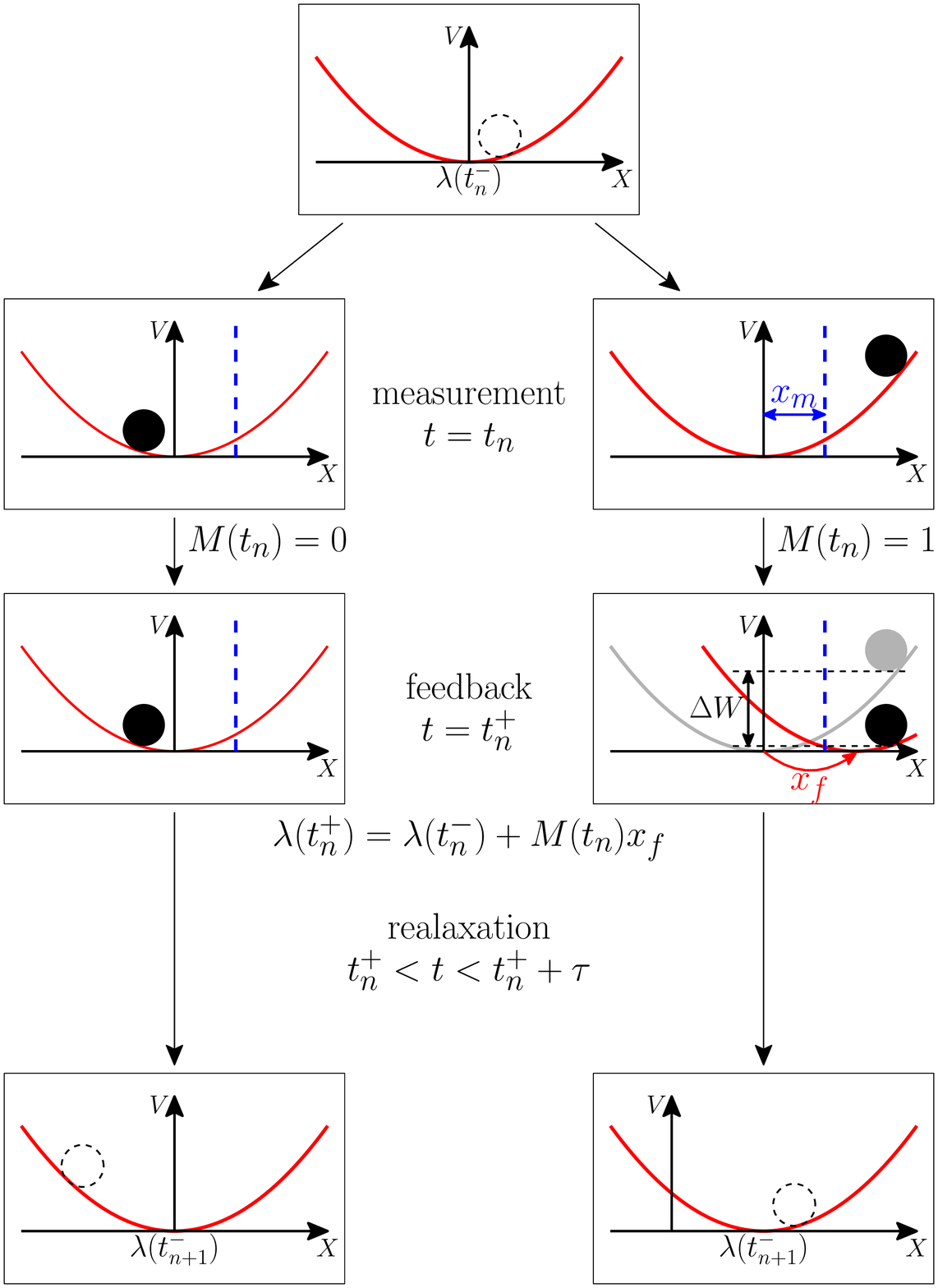}
\caption{(Color online) Illustration of the engine cycle during the time interval 
$t_n  \leq t < t_{n+1}$ of the engine.
At the measurement step, it is determined whether the particle is on the
left hand side~($M(t_n)=0$) or right hand side~($M(t_n)=1$) 
of the position $\lambda(t_n^-)+x_m$
represented by the (blue) dashed line. 
At the feedback step, if $M \left( t_n \right)=1$, the potential center is
instantaneously shifted to $\lambda(t_n^+) = \lambda(t_n^-)+x_f$ 
and the mechanical work $\Delta W (t_n)$ is extracted.
At the relaxation step, the particle is relaxed with the fixed potential
center at $\lambda(t_n^+)$ until the next cycle starts at time $t_{n+1}$.} 
\label{fig_op}
\end{figure}

The Brownian system can be used as an information engine by measuring $X$
and controlling $\lambda(t)$ depending on the measurement outcome.
Here, we consider the following time-periodic measurement and the
feedback control operations, which are also illustrated in Fig.~\ref{fig_op}.

%%  Fig. 1  %%%%%%%%%%%%%%%%%%%%%%%%%%%%%%%%%%%%%%%%%%
%%%%%%%%%%%%%%%%%%%%%%%%%%%%%%%%%%%%%%%%%%%%%%%%%%%%%%

{\it Measurement} -- 
At time $t = n \tau \equiv t_n$ $(n=0,1,2,...)$, a measurement is performed
to determine which side of a reference position at $\lambda(t_n)+x_m$ 
the Brownian particle is located at. 
The measurement outcome is represented by a binary parameter
\begin{equation}
M(t_n) = \left\{ \begin{array}{cl}
1 & \mbox{ if } X(t_n) \geq  \lambda(t_n)+x_m, \\ [2mm]
0 & \mbox{ if } X(t_n) < \lambda(t_n)+x_m  .
\end{array}\right.
\end{equation}
The information obtained during the measurement step can be exploited to
extract a work.

{\it Feedback control} --
When $M_n(t_n) = 0$, the potential center remains unchanged. That is,
\begin{equation}
\lambda(t_n^{+}) = \lambda(t_n^-),
\end{equation}
where $t_n^-~(t_n^+)$ denotes the moment just before~(after) the measurement
performed at time $t_n$. 
On the other hand, when $M(t_n)=1$, the potential center is shifted 
instantaneously by the amount of $x_f$: 
\begin{equation}
 \lambda(t_n^{+}) = \lambda(t_n^-) + x_f .
\end{equation}
By shifting the potential center, we can extract a work $\Delta W(t_n)$ 
as much as the change in the potential energy caused by the shift.
We adopt a convention that $\Delta W(t_n)$ is positive (negative) when 
the work is produced by (done on) the Brownian particle. 
It is given by
\begin{equation}\label{def_work}
\begin{split}
\Delta W(t_n) & =  V \left( X(t_n),\lambda(t_n^-) \right) - V \left(
X(t_n),\lambda(t_n^+) \right) \\
& =   x_f  \left(X(t_n)-\lambda(t_n^-) - \frac{1}{2}x_f \right).
\end{split} 
\end{equation}
Note that the extracted work is negative when $x_f<0$. 
Hence, we only consider the case with 
$x_f \geq 0$.

{\it Relaxation} --
In the time interval $t_n < t < t_{n+1}$, the particle evolves in time
with fixed $\lambda(t) = \lambda(t_n^+)$ 
according to the Langevin equation~\eqref{Langevin_eq} 
until the next cycle begins at time $t_{n+1}$.  
During this step, the particle exchanges the thermal energy with the heat
bath.

The engine is characterized by the three parameters:
$x_m$ for the measurement, $x_f$ for the feedback, and $\tau$ for the
relaxation. Thus, the extracted work per cycle or the power depends on the
choice of those parameters.
We are interested in the optimal choice of the parameters under which 
the steady state average of the extracted work per cycle or the power
becomes maximum. We remark that our model is a generalized version of the
information ratchet introduced in Ref.~\cite{Sagawa:2010th}, which
corresponds to the case with $x_f=2 x_m$ and $\tau=\infty$.

\section{Coordinate transformation}\label{sect_self}

The engine configuration is specified by the positions of the
particle $X$ and the potential center $\lambda$. Note that the potential center
is shifted by the amount of $x_f$ each time the measurement outcome is 1. 
Hence, it is convenient to
introduce an integer variable $l \equiv \lambda / x_f$ which counts the
number of potential-center shifts. We introduce $P_n(X,l)$ to
denote the joint probability distribution of $X$ and $l$ at time 
$t=t_n^-$.
The joint probability distribution satisfies the recursion relation
\begin{equation}\label{iter}
 \begin{split}
P_{n+1} \left( X,l \right) &=
\int_{-\infty}^{l x_f +x_m} K_{l x_f }^{\left( \tau \right)} 
\left( X | Z \right) P_n \left( Z,l \right)dZ  \\ 
& \hspace{-1cm} +\int_{(l-1)x_f+x_m}^{\infty}  K_{l x_f }^{\left( \tau \right)}
\left( X | Z \right) P_n \left( Z,l-1 \right) dZ ,
 \end{split}
\end{equation}
where
\begin{equation}\label{transition_prob}
K_{\alpha}^{(\tau)}(X|Z) = \frac{\exp\left[
-\frac{(X-\alpha-(Z-\alpha)e^{-\tau})^2}{2(1-e^{-2\tau})}\right]}
{\sqrt{2\pi (1-e^{-2\tau})}} 
\end{equation}
is the transition probability of the Brownian particle from position $Z$ at
time $0$ to position $X$ at time $\tau$ with the potential center being fixed
at position $\alpha$. Note that this is the transition probability for 
the Ornstein-Uhlenbeck process~\cite{Gardiner:2010tp, Risken:1996vl}.
The first~(second) term on the right hand side of Eq.~(\ref{iter}) accounts for 
the relaxation process after the feedback control corresponding to the
measurement outcome $M(t_n)=0~(1)$. 

\begin{figure}
\includegraphics*[width=\columnwidth]{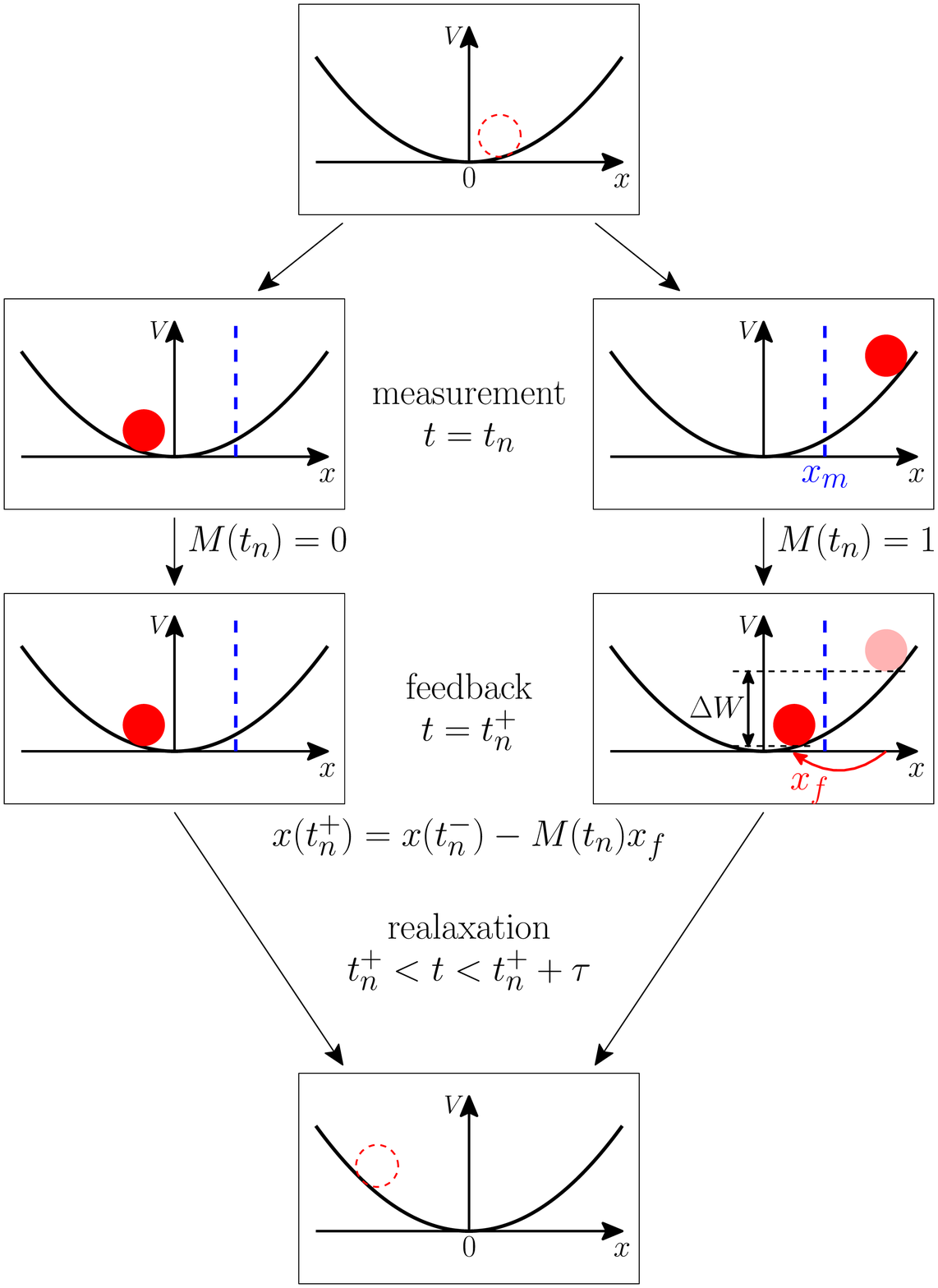}
\caption{(Color online) Illustration for the engine cycle 
in the fixed potential-center dynamics. In contrast to the original dynamics, 
the particle is transported instantaneously by the amount of $-x_f$ with
the potential center being fixed in the feedback process.}
\label{fig_op_co}
\end{figure}

The extracted work is determined only by the relative position of the
Brownian particle from the potential center.
Hence, it is useful to change the variables from $(X,l)$ to 
$(x\equiv X-lx_f,l)$.
%The variable $x$ measures the relative position of the Brownian
%particle from the potential center at $\lambda = lx_f$. 
Then, by using the translational invariance 
$K_\alpha^{(\tau)}(x+\alpha|z+\alpha)=K_0^{(\tau)}(x|z)$, we can
rewrite \eqref{iter} as
\begin{equation}\label{iter_indep_N}
 \begin{split}
P_{n+1} \left( x +lx_f ,l\right) = 
\int_{-\infty}^{x_m} K_0^{\left( \tau \right) } 
         \left( x | z \right) P_n \left( z+lx_f,l \right) dz\\
+ \int_{x_m}^{\infty} K_0^{\left( \tau \right) } \left( x | z-x_f \right)
P_n \left( z+(l-1)x_f,l-1 \right) dz. 
 \end{split}
\end{equation}
By summing Eq.~(\ref{iter_indep_N}) over all $l$, we obtain
% and defining the new probability distribution $p_n(x)$ as
%we obtain the $l$-independent recursive relation:
\begin{equation}\label{iter_cm}
 \begin{split}
p_{n+1} \left( x \right) &= \int_{-\infty}^{x_m} K_0^{\left( \tau \right) }
\left( x | z \right) p_n \left( z \right) dz\\
&+ \int_{x_m}^{\infty} K_0^{\left( \tau \right) } \left( x | z-x_f
\right) p_n \left( z \right) dz , 
 \end{split}
\end{equation}
where 
\begin{equation}
p_n \left( x \right) \equiv \sum_{l} P_n \left( x+lx_f,l \right).
\end{equation}
is the probability distribution function for the relative position $x$ at
time $t_n^-$.
This recursion relation can be understood in terms of an effective dynamics. 
In the effective dynamics, the potential center is fixed at the origin.
Instead, the Brown particle is instantaneously shifted by the amount of 
$(-x_f)$ when the measurement outcome is $M=1$.
This effective dynamics is illustrated in
Fig.~\ref{fig_op_co} and will be referred to as the `fixed potential-center
dynamics'.

%%  Fig. 2  %%%%%%%%%%%%%%%%%%%%%%%%%%%%%%%%%%%%%%%%%%
%%%%%%%%%%%%%%%%%%%%%%%%%%%%%%%%%%%%%%%%%%%%%%%%%%%%%%

In the $n\rightarrow \infty$ limit, $p_n(x)$ will
converge to the steady-state distribution $p_{ss}(x)$, 
which is given by the solution of the self-consistent equation 
\begin{equation}\label{self_pss}
 \begin{split}
p_{ss} \left( x \right)
&= \int_{-\infty}^{x_m} K_0^{\left( \tau \right)} \left( x|z \right) p_{ss}
\left( z \right) dz\\
&+\int_{x_m}^{\infty} K_0^{\left( \tau \right)} \left( x|z-x_f \right)
p_{ss} \left( z \right) dz.
 \end{split}
\end{equation}
%We will calculate the average extracted work in the steady state limit. 
From Eq.~(\ref{def_work}), the work is extracted only when $x>x_m$ by the
amount of $\Delta W = x_f(x-x_f/2)$ each cycle. 
Hence, the average extracted work per cycle in the steady state is given by
\begin{equation}\label{W_ss}
\langle \Delta W\rangle_{ss} 
= x_f \int_{x_m}^{\infty} \left(x-\frac{1}{2}x_f\right) p_{ss}(x) dx ,
\end{equation}
where $\langle \cdots \rangle_{ss}$ denotes the steady-state ensemble average. 
The integration in Eq.~\eqref{W_ss} begins at $x_m$ because the work can be
extracted only when the particle position is larger than $x_m$~($M=1$). 
Such an event occurs with the probability $P_M$ given by  
\begin{equation}
P_M \equiv \int_{x_m}^\infty \ p_{ss}(x) dx \ .
\end{equation}
Using this quantity, we can write $ \langle \Delta W\rangle_{ss}$ as
\begin{equation}\label{Wss_PM}
\langle \Delta W\rangle_{ss} =  x_f \left( \langle x\rangle_M -\frac{1}{2}
x_f\right) P_M ,
\end{equation}
where
\begin{equation}
\langle x\rangle_M \equiv \frac{1}{P_M} \int_{x_m}^\infty x\ {p_{ss}(x)} dx
\end{equation}
is the mean position of the particle in the steady state given that $x\geq
x_m$. The system acts as an engine with positive $\langle \Delta
W\rangle_{ss}$ when
\begin{equation}
0 < x_f < 2 \langle x\rangle_M \ .
\end{equation}

\section{Optimal condition for the engine}\label{sect_result}

In this section, we develop an analytic formalism for $p_{ss}(x)$ and 
discuss the optimal operating condition for the engine. 
We address the special cases in the limit 
$\tau\to\infty$ and $\tau\to 0$, then proceed to 
the general case with nonzero and finite $\tau$.

\subsection{$\tau \rightarrow \infty$ case}

When $\tau$ is infinite, the system relaxes to the equilibrium
state irrespective of the measurement and the feedback control. 
Thus, the system follows the equilibrium distribution
\begin{equation}\label{K_0_infty}
p_{ss}(x) = \frac{1}{\sqrt{2\pi}} e^{-\frac{1}{2}x^2}.
\end{equation}
It is easy to check that the equilibrium distribution is indeed the solution
of the self-consistent equation~\eqref{self_pss} with $K_0^{\infty}(x|z) =
e^{-x^2/2}/\sqrt{2\pi}$.
Using this $p_{ss}(x)$, we obtain that
\begin{equation}
P_M = \frac{1}{2}\erfc\left(\frac{x_m}{\sqrt{2}} \right) \mbox{ and }
\langle x\rangle_M = \sqrt{\frac{2}{\pi}}
\frac{e^{-\frac{1}{2}x_m^2}}{\erfc\left(\frac{x_m}{\sqrt{2}}\right)} \label{P_M_x_M}
\end{equation}
where $\erfc(x) = \frac{2}{\sqrt{\pi}}\int_x^\infty e^{-y^2}dy$
is the complementary error function.

The optimal values of $x_f$ and $x_m$ at which the engine extracts the
maximum amount of works are denoted by $x_f^*$ and $x_m^*$, respectively. 
They are obtained from the conditions
$\partial\langle \Delta W\rangle_{ss}/\partial x_f |_{x_f=x_f^*,x_m=x_m^*}=0$ 
and $\partial\langle \Delta W\rangle_{ss}/\partial x_m 
|_{x_f=x_f^*,x_m=x_m^*}=0$, which yield that
\begin{eqnarray}
x_f^* &=& \langle x\rangle_M |_{x_m = x_m^*}, \label{opt_x_f} \\ 
x_m^* &=& \frac{x_f^*}{2}. \label{opt_x_m}
\end{eqnarray}
The former equation~\eqref{opt_x_f} for $x_f^*$ has a clear meaning: 
Given a particle position $x>x_m$, the work is extracted maximally 
by shifting the Brownian particle to the potential 
center~(in the fixed potential-center dynamics).
Thus, $x_f^*$ should be taken as the mean position of the particle under the
condition that $M=1$.
With the optimal choice of $x_f$, the mean value
of the extracted work is
given by $\langle\Delta W\rangle_{ss} = \frac{1}{2}\langle x\rangle_M^2
P_M$. Note that $\frac{1}{2}\langle x\rangle_M^2$ is an increasing function
of $x_m$ while $P_M$ is a decreasing function of $x_m$~(see \eqref{P_M_x_M}).
%This leads to a competition: As one increases $x_m$, the extracted 
%work~($\langle x\rangle_M^2/2$) per each feedback with $M=1$ increases 
%while the probability of such an event decreases. 
Due to these competing effects, the work becomes
maximum at a nontrivial value of $x_m^*$. Combining \eqref{opt_x_f} and 
\eqref{opt_x_m}, one obtains the transcendental equation for $x_m^*$:
\begin{equation}
x_m^* = \frac{1}{\sqrt{2\pi}}
\frac{e^{-\frac{1}{2}x_m^*}}{\erfc(x_m^*/\sqrt{2})}. 
\end{equation} 
It has the numerical solution $x_m^* \simeq 0.612$. Therefore, $x_f^* =
2x_m^* \simeq 1.224$ and the maximum average work per cycle 
$\left< \Delta W \right>_{ss}^* $ is given by
\begin{equation}
\left< \Delta W \right>_{ss}^* 
= \left. \frac{1}{2} \langle x\rangle_M^2 P_M \right|_{x_m=x_m^*} \simeq
0.202.
\end{equation}

\begin{figure*}
\includegraphics[width=0.66\columnwidth]{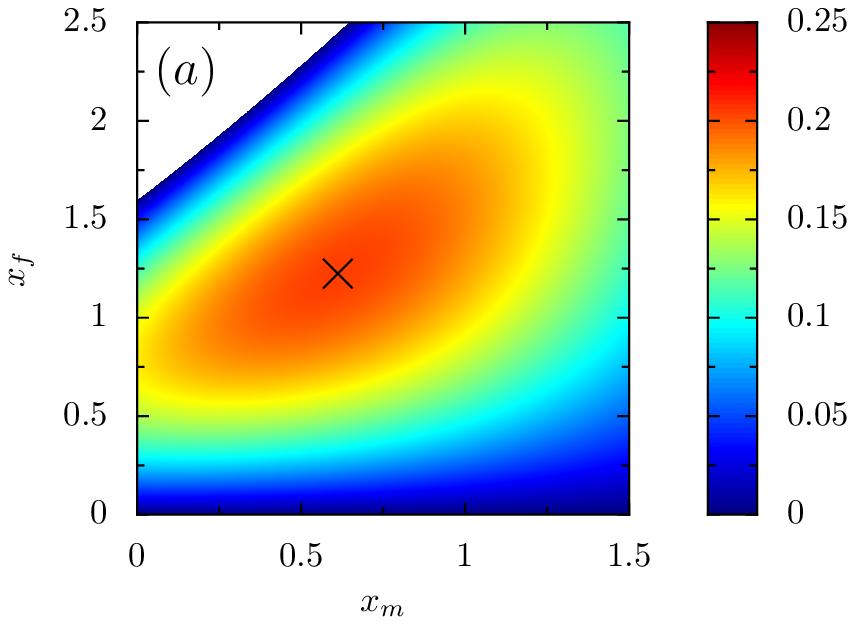}
\includegraphics[width=0.66\columnwidth]{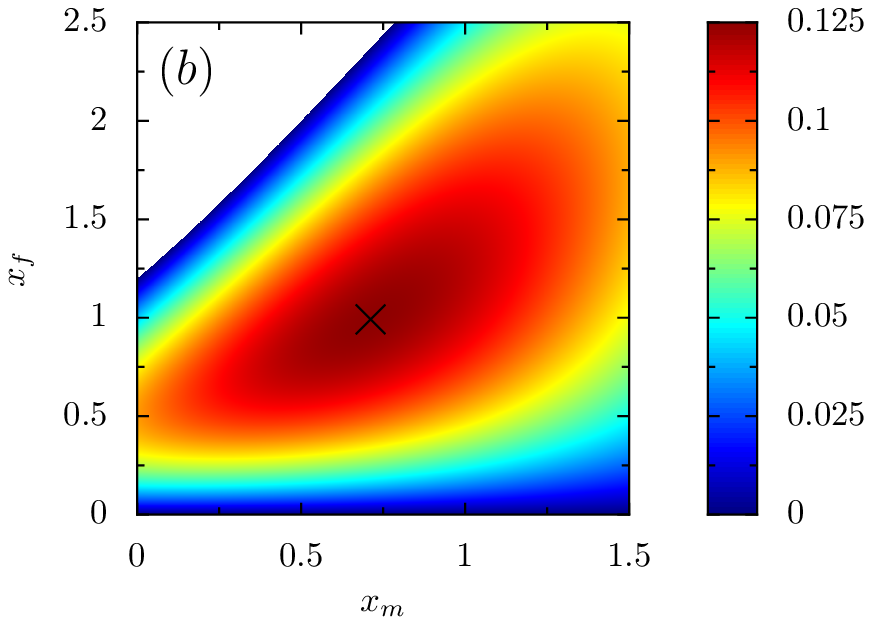}
\includegraphics[width=0.66\columnwidth]{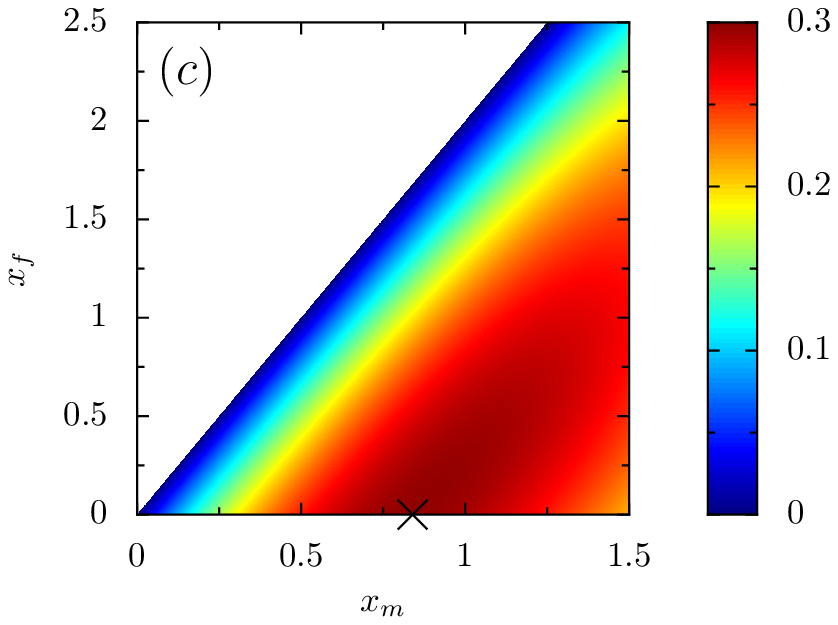}
\caption{(Color online) (a) Density plot for $\langle \Delta W \rangle_{ss}$ 
in $\tau \rightarrow \infty$ limit.
(b) Density plot for $\langle \Delta W \rangle_{ss}$ 
obtained from the truncation method when $L=4$ and $\tau=\log 2$.
(c) Density plot for $w_{ss}$ in $\tau \rightarrow 0$
limit. The $times$ symbols represent the optimal position where the
extracted work or the power is maximum.}
\label{fig_densityplot}
\end{figure*}

Figure~\ref{fig_densityplot}(a) shows the density plot of the average
extracted work per cycle in the $(x_m,x_f)$ plane. 
The work is indeed maximum at $(0.612, 1.224)$.
We add a remark that the average power $\langle \Delta W\rangle_{ss}^*$ 
is zero because 
$\left< \Delta W \right>_{ss}^*$ is finite but $\tau \rightarrow \infty$.

%%  Fig. 3  %%%%%%%%%%%%%%%%%%%%%%%%%%%%%%%%%%%%%%%%%%

%%%%%%%%%%%%%%%%%%%%%%%%%%%%%%%%%%%%%%%%%%%%%%%%%%%%%%
\subsection{$\tau \rightarrow 0$ case}

In the $\tau \rightarrow 0$ limit, the particle position is measured
incessantly.
Therefore, in the fixed potential-center dynamics, the particle is
immediately shifted from $x_m$ to $x_r \equiv x_m - x_f$ whenever it touches the reference position $x_m$. This dynamics is similar to the resetting process studied by Evans and
Majumdar~\cite{Evans:2011jo}. They investigated a search problem by 
a random walker whose position is reset to the origin at a constant rate.
Along the similar line of reasoning, our resetting process can be described 
by the following Fokker-Planck equation:
\begin{equation}
\frac{\partial p(x,t)}{\partial t} = - \frac{\partial j(x,t)}{\partial x} +
j_r(t) \delta(x-x_r)
\end{equation}
where $p(x,t)$ is the probability distribution of the particle in the fixed
potential-center dynamics,  
\begin{equation}
j(x,t) = \left( -x -\frac{\partial}{\partial x}\right) p(x,t)
\end{equation}
is the probability current at position $x$ and at time $t$, and 
$j_r(t) = \lim_{x\to x_m^-}j(x,t)$ is the resetting current which is 
absorbed at $x_m$ and then injected at $x_r$.
The probability distribution satisfies the absorbing boundary
condition at $x_m$, i.e., $p(x=x_m,t)=0$.

The steady-state probability distribution satisfies 
\begin{equation}\label{current_def}
-x p_{ss}(x) - \frac{\partial p_{ss}(x)}{\partial x} =  
\begin{cases}
0 & \mbox{for }x<x_r   \\
j_{ss} & \mbox{for }x_r\leq x<x_m  \\
0 & \mbox{for } x\geq x_m
\end{cases}
\end{equation}
where the steady-state resetting current is given by
\begin{equation}
j_{ss} = \lim_{t\to\infty} j_r(t) = -\left.\frac{\partial
p_{ss}(x)}{\partial x}\right|_{x=x_m^-}.
\end{equation} 
%which has to be determined self-consistently.
Given $j_{ss}$, the solution satisfying the absorbing 
boundary condition is given by 
\begin{equation}
p_{ss} \left( x \right)=
\begin{cases}
j_{ss} \int_{x_r}^{x_m}e^{\frac{1}{2}
(z^2-x^2)}dz & \mbox{for }x<x_r ,\\
 j_{ss} \int_x^{x_m}e^{\frac{1}{2} (z^2-x^2) }dz &
\mbox{for }x_r\leq x<x_m ,\\
0 & \mbox{for }x \geq x_m .
\end{cases}
\end{equation}
The resetting current is determined by the normalization condition 
$\int p_{ss}(x)dx=1$. It is given by
\begin{equation}
\begin{split}
j_{ss} = & \left[
\left( \int_{-\infty}^{x_r}e^{-\frac{1}{2} x^2 } dx\right)
\left(\int_{x_r}^{x_m}e^{\frac{1}{2} z^2 }dz\right) \right.\\
& +\left.\int_{x_r}^{x_m}\left( e^{-\frac{1}{2} x^2 }
\int_x^{x_m}e^{\frac{1}{2} z^2}dz \right)dx
\right]^{-1} \label{j_ss}
\end{split}
\end{equation}
with $x_r = x_m - x_f$.

In the $\tau\to 0$ limit, the average extracted work per cycle vanishes
because it takes infinitely many cycles for the Brownian particle to 
reach $x_m$ after a resetting. 
Thus, it is useful to consider the average power in the steady state 
$w_{ss} \equiv \lim_{\tau\to 0}\langle \Delta W\rangle_{ss} /\tau$. 
It is given by 
\begin{equation}
w_{ss} = x_f \left( x_m - \frac{1}{2}x_f \right) j_{ss},
\end{equation}
where $x_f (x_m - x_f/2) = V(x_m,0)-V(x_m-x_f,0)$ is the extracted work
per resetting. 
Figure \ref{fig_densityplot}(c) shows the density plot for $w_{ss}$ in the
$\tau\to 0$ limit in the $(x_m, x_f)$ plane. 

The power is maximized when $\partial w_{ss}/\partial x_m = 0$ and $\partial
w_{ss}/\partial x_f=0$ simultaneously. A straightforward calculation shows
that both conditions become identical when $x_f=0$, which implies that
$x_f^*=0$. In the limit $x_f \to x_f^*=0$, the power becomes 
$w_{ss} = \sqrt{\frac{2}{\pi}} x_m e^{-x_m^2/2} / 
(1+\text{erf}(x_m/\sqrt{2}))$ with the error function ${\rm erf}(x) = 1-{\rm
erfc}(x)$. It takes the maximum value
\begin{equation}
w_{ss}^* \simeq 0.295 ~\textrm{at } x_m^* \simeq 0.840 \textrm{ and } x_f^* = 0. \label{max_power_at_zero}
\end{equation}
Strictly speaking, the engine does not produce any work at $x_f=0$.
The result $x_f^*=0$ should be understood as the limit $x_f \to
0^+$. In this limit, $V(x_m,0)-V(x_m-x_f,0)$, the work extracted in a
feedback process, vanishes as $\mathcal{O}(x_f)$, 
but the resetting current $j_{ss}$ in
\eqref{j_ss} diverges as $\mathcal{O}(x_f^{-1})$, 
which results in a finite power.

\subsection{Finite $\tau$ case}

For finite $\tau$, $p_{ss}(x)$ cannot be obtained in a closed form. 
Thus, we try to find it in a series form 
\begin{equation}\label{series}
p_{ss} (x)=\sum_{n=0}^\infty c_n \phi_n(x)
\end{equation}
using the basis functions 
$\phi_n(x) \equiv H_n\left(x/\sqrt{2}\right)e^{-\frac{1}{2}x^2}$ 
where $H_n \left( x \right)$ is the Hermite polynomial of degree
$n$~\cite{Arfken:2013uf}.
The Hermite polynomials satisfy the orthogonality condition
\begin{equation}\label{normal}
\int_{-\infty}^{\infty} H_n\left(\frac{x}{\sqrt{2}}\right)
                        H_l\left(\frac{x}{\sqrt{2}}\right)
e^{-\frac{1}{2}x^2} dx =  N_n {\delta_{nl}}  
\end{equation}
with $N_n \equiv \sqrt{2\pi}2^n n!$. 
The expansion coefficients are represented by a column vector 
$\bm{c} = (c_0,c_1,c_2,\ldots)^T$ where the superscript ${}^T$ stands for 
the transpose. 
The normalization condition $\int p_{ss}(x)dx=1$ fixes $c_0=1/\sqrt{2\pi}$. 
The other coefficients will be determined by using the self-consistent 
equation \eqref{self_pss}.

Such an expansion \eqref{series} is natural because 
$\phi_n(x)$ is the eigenfunction of the Fokker-Planck operator 
$\mathcal{L}(x) = \frac{\partial}{\partial
x}\left(x+\frac{\partial}{\partial x}\right)$ for the
Ornstein-Uhlenbeck process~\cite{Risken:1996vl}, i.e., 
\begin{equation}
\mathcal{L}(x) \phi_n(x) = -n \phi_n(x)  .
\end{equation}
The transition probability $K_0^{(\tau)}(x|z)$ in Eq.~\eqref{self_pss} 
can be written in terms of $\mathcal{L}(x)$ as 
$K_0^{(\tau)}(x|z) = e^{\tau \mathcal{L}(x)} \delta(x-z)$.
Thus, Eq.~(\ref{self_pss}) can be rewritten as
\begin{equation}\label{self_pss_L}
p_{ss} \left( x \right)=e^{\tau\mathcal L(x)} \left[
p_{ss}^0(x) + p_{ss}^1(x+x_f) \right] ,
\end{equation} 
where $p_{ss}^0(x) \equiv \Theta(x_m-x) p_{ss}(x)$ and $p_{ss}^1(x) \equiv 
 \Theta(x-x_m) p_{ss}(x) = p_{ss}(x) - p_{ss}^0(x)$
with the Heaviside step function $\Theta(x)$. 

Our strategy is to expand both sides of \eqref{self_pss_L} 
using the basis set $\{\phi_n\}$. 
First of all, the function $p_{ss}^0(x) + p_{ss}^1(x+x_f)$ in the right hand 
is expanded as
\begin{equation}\label{p'_series}
p_{ss}^0(x) + p_{ss}^1(x+x_f) = \sum_{n=0}^\infty c'_n \phi_n(x)  .
\end{equation}
The expansion coefficients $\bm{c}' = (c_0',c_1',\cdots)^T$ are obtained by
integrating both sides of \eqref{p'_series} after being multiplied with
$H_m(x/\sqrt{2})$. One obtains that 
\begin{equation}
\bm{c}' = (\mathsf{A} + \mathsf{B} ) \bm{c}, \label{c'}
\end{equation}
where the matrix elements of $\mathsf{A}$ and $\mathsf{B}$ are defined as
\begin{eqnarray}
A_{nl} &=& \frac{1}{N_n} \int_{-\infty}^{x_m}
           H_n\left(\frac{x}{\sqrt{2}}\right) \phi_l(x)dx, \label{A_elements}\\
B_{nl} &=& \frac{1}{N_n} \int_{x_m}^\infty 
           H_n\left(\frac{x-x_f}{\sqrt{2}} \right) \phi_l(x)dx  .\label{B_elements}
\end{eqnarray}
Note that the Hermite polynomials satisfy the identity
\begin{equation}\label{H_identity}
H_n(x+y) = H_n(x) + \sum_{k=0}^{n-1} \binom{n}{k} (2y)^{n-k}  H_k(x),
\end{equation}
with the binomial coefficient $\binom{n}{k}$. This identity allows us to
rewrite $\textsf{A+B}$ as
\begin{equation}
\mathsf{A+B} = \mathsf{I+F} , \label{I+F}
\end{equation}
where $\mathsf{I}$ is the identity matrix and $\mathsf{F}$ has the 
elements
$$
F_{nl} = \frac{1}{N_n} \sum_{k=0}^{n-1} \binom{n}{k} 
\left(-\sqrt{2}x_f\right)^{n-k} 
 \int_{x_m}^\infty H_k\left(\frac{x}{\sqrt{2}}\right) \phi_l(x) dx 
$$
for $n\geq 1$ and $F_{nl} = 0$ for $n=0$.
Using $e^{\tau\mathcal{L}(x)} \phi_n(x) =
e^{-n\tau}\phi_n(x)$ and introducing a diagonal matrix $\textsf{W}$ with
elements $W_{nl} = e^{-n\tau}\delta_{nl}$, we finally obtain the self-consistent equation
\begin{equation}\label{self_c}
\bm{c} = \mathsf{W} (\mathsf{I+F}) \bm{c}. 
\end{equation}
It is more convenient to work with 
\begin{equation}
\bm{d} = \mathsf{W}^{-1} \bm{c} ,
\end{equation}
with which the self-consistent equation~\eqref{self_c} becomes
\begin{equation}\label{self_d}
\bm{d} = (\mathsf{I+F})\mathsf{W} \bm{d}  .
\end{equation}

The average extracted work per cycle is given by
\begin{equation}
\begin{split}
\langle \Delta W\rangle_{ss} =& \int_{x_m}^\infty \left(\frac{1}{2}x^2
-\frac{1}{2}(x-x_f)^2 \right) p_{ss}(x)dx \\
= & \int_{-\infty}^\infty \frac{1}{2}x^2 \left(
p_{ss}^1(x) - p_{ss}^1(x+x_f) \right) dx .
\end{split}
\end{equation}
The involved distribution functions are expanded as 
$(p_{ss}^1(x)-p_{ss}^1(x+x_f)) = -\sum_n (\mathsf{F} \bm{c})_n\phi_n(x)$
using \eqref{p'_series}, \eqref{c'}, and \eqref{I+F}. 
Note that $x^2 = H_2(x/\sqrt{2})/2+H_0(x/\sqrt{2})$. Hence, using the
orthogonality~\eqref{normal} of the Hermite polynomials, 
we obtain that
\begin{equation}\label{W2}
\langle \Delta W\rangle_{ss} = -\sqrt{8\pi} (\mathsf{F} \bm{c})_2 
= -\sqrt{8\pi}\left(1-e^{-2\tau}\right) d_2 \ .
\end{equation}
For the second equality, $\mathsf{F}\bm{c} =
(\mathsf{W}^{-1}-\mathsf{I})\bm{c}=(\mathsf{I}-\mathsf{W})\bm{d}$ is used.

The formal solution of $\bm{d}$ is easily derived. First, we write
$\bm{d}$, $\mathsf{W}$, and $\mathsf{F}$ in a block form as 
\begin{equation}
\bm{d} = (d_0,\tilde{\bm{d}})^T,\
\mathsf{W} = \left( \begin{array}{cc}
      1 & 0  \\
      0 & \widetilde{\mathsf{W}} \end{array}\right) \mbox{, and }
\mathsf{F} = \left( \begin{array}{cc}
   0 & 0 \\
   \tilde{\bm f} & \widetilde{\mathsf{F}} \end{array}\right) ,
\end{equation}
and define the column vectors $\tilde{\bm{d}} = (d_1,d_2,\cdots)^T$ with 
$d_n=e^{n\tau}c_n$ and $\tilde{\bm{f}}=(f_1,f_2,\cdots)^T$ with $f_n =
F_{n0}$,  and matrices $\widetilde{\mathsf{W}}$ and 
$\widetilde{\mathsf{F}}$ accordingly. 
%Due to the normalization of the probability distribution, 
%where $d_0 = c_0 = 1/\sqrt{2\pi}$ and $\tilde{\bm f} = (f_1,f_2,\ldots)^T$
%with $f_n = F_{n0}$~($n=1,2,\cdots$). 
Inserting these block forms into \eqref{self_d}, we obtain the formal
solution for $\tilde{\bm{d}}$ as
\begin{equation}\label{solution}
\tilde{\bm{d}} = c_0 \left[ \mathsf{I} -
(\mathsf{I}+\widetilde{\mathsf{F}})\widetilde{\mathsf{W}}\right]^{-1}
\tilde{\bm{f}}
\end{equation}
with $c_0 = 1/\sqrt{2\pi}$.
It is crucial to have the formal solution that the first row of $\mathsf{F}$
vanishes~($F_{0n}=0$).

The formal solution involves an inversion of infinite-dimensional matrices,
hence a closed-form expression is not available.
Nevertheless, it is useful because it enables us to obtain an approximate
solution systematically. 
First, we truncate the matrices 
$\widetilde{\mathsf{F}}$ and $\widetilde{\mathsf{W}}$ to 
$L\times L$ matrices $\widetilde{\mathsf{F}}^{(L)}$ and 
$\widetilde{\mathsf{W}}^{(L)}$ and 
the vector $\widetilde{\bm{f}}$ to an $L\times
1$ vector $\tilde{\bm{f}}^{(L)}$, respectively, 
i.e., $\widetilde{F}^{(L)}_{nm}=\widetilde{F}_{nm}$,
$\widetilde{W}^{(L)}_{nm}=\widetilde{W}_{nm}$, and $\tilde{f}^{(L)}_n =
\tilde{f}_n$ for $n,m = 1,2,\dots,L$. They are inserted into 
\eqref{solution} to yield a truncated solution $\tilde{\bm{d}}^{(L)}$.
We note that $\widetilde{\mathsf{F}}$ and $\tilde{\bm{f}}$ depend on $x_m$ and
$x_f$ but not on $\tau$, while the diagonal matrix 
$\widetilde{\mathsf{W}}$ depends only on $\epsilon \equiv e^{-\tau}\leq 1$.
Therefore, the truncated solution $\tilde{\bm{d}}^{(L)}$ is the exact up to
$\mathcal{O}(\epsilon^L)$. Then, from Eq.~\eqref{W2}, we can obtain the
approximate solution for the average extracted work $\langle \Delta 
W\rangle_{ss}^{(L)}= -\sqrt{8\pi}\left(1-e^{-2\tau}\right)
\tilde{d}_2^{(L)}$, which is also exact up to $\mathcal{O}(\epsilon^L)$.

\begin{figure}
\includegraphics*[width=\columnwidth]{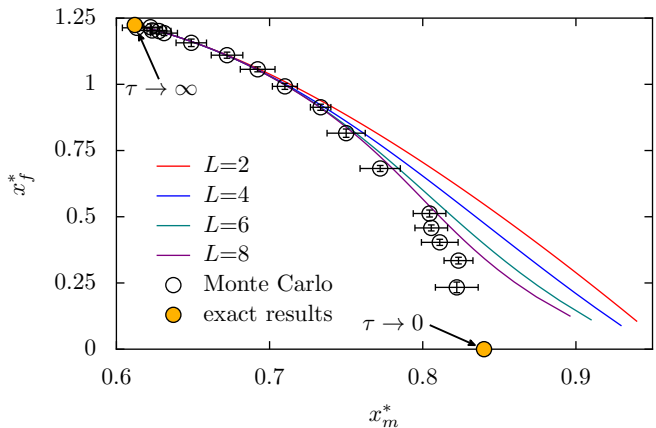}
\caption{Parametric plot for the optimal control parameters 
$(x_m^*(\tau),x_f^*(\tau))$. The solid lines are obtained from the truncation method for $L=2,4,6,8$. The results obtained from the Monte Carlo
simulations are denoted by open circles with error bars. The exact results in $\tau \rightarrow 0$ and $\tau \rightarrow \infty$ are marked by arrows.}
\label{fig_optimalconditions}
\end{figure}

Figure~\ref{fig_densityplot}(b) shows the density plot for 
$\langle \Delta W\rangle_{ss}^{(L)}$ with $L=4$ and at $\tau=\ln 2$. 
It is maximum at the point $(x_m^*,x_f^*)$ marked by the symbol 
$\times$ whose position can be found numerically. 
%In this way, we can find the optimal values $x_m^*(\tau)$ and $x_f^*(\tau)$
%leading to the maximum work at each value of $\tau$ and $L$.
In Fig.~\ref{fig_optimalconditions}, we present the traces of 
$(x_m^*,x_f^*)$ as $\tau$ is varying with $L=2,4,6,8$. 
Along each line, $x_m^*~(x_f^*)$ decreases~(increases) as $\tau$ increases. 
The lines converge to a single curve at large values of $\tau$. 
The convergence becomes poor in the region of small $\tau$ 
where the truncation parameter $\epsilon=e^{-\tau}$ is not small.

%%  Fig. 4  %%%%%%%%%%%%%%%%%%%%%%%%%%%%%%%%%%%%%%%%%%
%%%%%%%%%%%%%%%%%%%%%%%%%%%%%%%%%%%%%%%%%%%%%%%%%%%%%%

We also performed the Monte Carlo simulations to obtain the optimal
parameter values.
In the Monte Carlo simulations, the Langevin equation~\eqref{Langevin_eq}
was integrated numerically over $10^9$ engine cycles to estimate 
the average extracted work in the steady state. In order to estimate 
$x_m^*$ and $x_f^*$, we discretized $x_m$ and $x_f$ in units
of $\Delta x_m = \Delta x_f = 0.01$.
Among the grid points of $(x_m,x_f)$, we selected nine points having the
largest values of $\langle \Delta W\rangle_{ss}$. 
Their averages were taken as the 
the Monte Carlo results and the standard deviations as the error bars for
$x_m^*$, $x_f^*$, and $\langle \Delta W\rangle_{ss}^*$.
The simulated $x_m^*$ and $x_f^*$ are plotted with open symbols 
in Fig.~\ref{fig_optimalconditions} with error bars.
The exact optimal values in the $\tau \rightarrow 0$ and $\tau \rightarrow
\infty$ limits are also plotted in Fig.~\ref{fig_optimalconditions} with
closed symbols for comparison. As seen in the figure, our simulated data at
large and small $\tau$ are close to the exact results in $\tau \rightarrow
\infty$ and $\tau \rightarrow 0$, respectively, which supports the validity of our Monte Carlo simulations.
The analytic results are in good agreement with the Monte Carlo results
unless $\tau$ is too small.

\begin{figure}
\includegraphics*[width=\columnwidth]{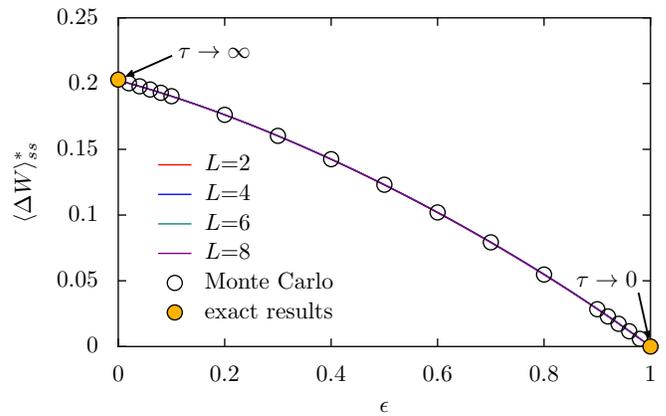}
\caption{Parametric plot for the optimal work per cycle $\langle \Delta W\rangle_{ss}^*$. Results obtained from the truncation method are denoted by solid lines. Open circles present the Monte Carlo
simulations results. The exact results in $\tau \rightarrow 0$ and $\tau \rightarrow \infty$ are marked by arrows.} 
\label{fig_work}
\end{figure}

Figure~\ref{fig_work} presents the plot of $\langle \Delta W\rangle_{ss}^*$
as a function of $\epsilon=e^{-\tau}$. As the figure shows, the Monte Carlo
results~(open symbols) and the analytic results~(lines) agree perfectly
well even for small values of $L$. The numerical results show that $\langle
\Delta W\rangle_{ss}^*$ increases as $\tau$ increases so that  
the global maximum of $\langle \Delta W\rangle_{ss}^*$ is attained
when $\tau \rightarrow \infty$. 
Note the extracted work from an information engine is bounded by
the change in the mutual information between the engine and the
measurement outcome during the relaxation 
process~\cite{{Sagawa:2010th},{Sagawa:2012th2}}.
When $\tau \rightarrow \infty$, the mutual information generated at 
the measurement step completely vanishes during the relaxation step. 
This might be the reason why the average extracted work is maximum at 
$\tau \rightarrow \infty$.

%%  Fig. 5  %%%%%%%%%%%%%%%%%%%%%%%%%%%%%%%%%%%%%%%%%%
%%%%%%%%%%%%%%%%%%%%%%%%%%%%%%%%%%%%%%%%%%%%%%%%%%%%%%
%%  Fig. 6  %%%%%%%%%%%%%%%%%%%%%%%%%%%%%%%%%%%%%%%%%%
\begin{figure}
\includegraphics*[width=\columnwidth]{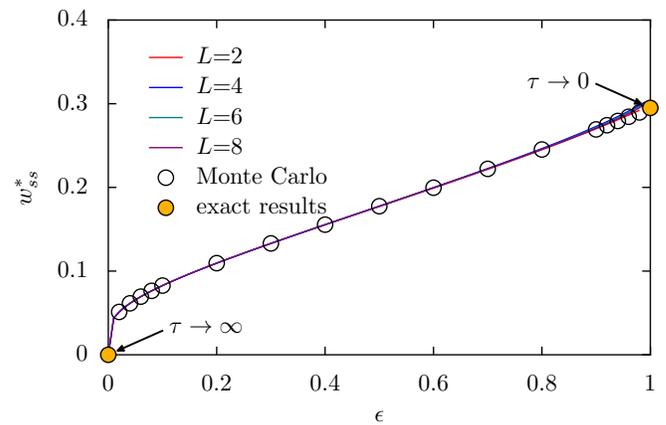}
\caption{Parametric plot for the optimal power $w_{ss}^*$. Results obtained from the truncation method are denoted by solid lines. Open circles present the Monte Carlo
simulations results. The exact results in $\tau \rightarrow 0$ and $\tau \rightarrow \infty$ are marked by arrows.} 
\label{fig_power}
\end{figure}
%%%%%%%%%%%%%%%%%%%%%%%%%%%%%%%%%%%%%%%%%%%%%%%%%%%%%%

Figure~\ref{fig_power} shows the optimal power $w_{ss}^*=\langle \Delta
W\rangle_{ss}^*/\tau$ as a function of $\epsilon$. 
In contrast to the optimal work $\langle \Delta W\rangle_{ss}^*$ per cycle, 
the optimal power $w_{ss}^*$ is
a decreasing function of $\tau$ and becomes maximum in the limiting case 
$\tau \to 0$. This indicates that the continuous time operation is the best
way to achieve the maximum power of the Brownian information engine. Our
model assumes that measurement and feedback processes do not cost any
energy. If they cost some energy, the global maximum of the power will be
realized at finite $\tau$. 

\section{Conclusion}\label{sect_conclusion}
We studied the information engine where the Brownian particle is confined in
a harmonic potential. This engine consists of the three processes:
measurement of the particle position, instantaneous shift of the potential
center depending on the measurement outcome, and relaxation of the particle.
Each process is characterized by the model parameter: $x_m$ for the
measurement, $x_f$ for the feedback, and $\tau$ for the relaxation. Using
the coordinate transformation, we derived the self-consistent equation for
the steady state distribution function of the particle in the fixed
potential-center dynamics. The average work extracted out of the information 
engine per cycle is found from the steady state distribution.

When $\tau \rightarrow \infty$, the steady state becomes the equilibrium
state. When $\tau\to 0$, the dynamics becomes similar to the resetting
process~\cite{Evans:2011jo} and the exact steady state distribution is
obtained by analyzing the corresponding Fokker-Planck equation.
When $\tau$ is finite, the steady-state distribution has the infinite series
expansion in terms of the Hermite polynomials, which can be approximated
systematically by truncating the infinite series. We show that the extracted
work per cycle is maximum at $\tau=\infty$ and the the extracted power is
maximum in the limiting case $\tau\to 0$.

A Brownian particle confined by a harmonic potential is realized by the optical
trap experiment as in e.g. Ref.~\cite{Lee:2015hn}. We expect that our
theoretical model can be tested in such experiments. In our model, the
Brownian particle exhibits a ballistic motion as the engine operates. This
suggests that one can design an information motor which rectifies the
thermal fluctuations with the help of measurement and feedback controls.
Further studies along this direction would be interesting.

\begin{acknowledgments}
This research was supported by the National Research Foundation (NRF) of
Korea Grant Nos. 2013R1A2A2A05006776~(JDN) and ~2011-35B-C00014 (JSL).
\end{acknowledgments}

\appendix
\bibliographystyle{apsrev}
\bibliography{paper}

\end{document}